\documentclass[final,a4paper,notitlepage,twocolumn]{article}
\usepackage{amsmath}

\newtheorem{theorem}{Theorem}
\newtheorem{acknowledgement}[theorem]{Acknowledgement}

\input{tcilatex}

\begin{document}

\title{The Loss of fidelity due to quantum leakage\\
for Josephson charge qubits}
\author{Xian-Ting Liang$^{\thanks{%
Email address: xtliang@ustc.edu}}$, Yong-Jian Xiong \\
%EndAName
Department of Physics and Institute of Modern Physics,\\
Ningbo University, Ningbo, Zhejiang 315211, China}
\maketitle

\begin{abstract}
In this paper we calculate the loss of fidelity due to quantum leakage for
the Josephson charge qubit (JCQ) in virtue of the Mathieu functions. It is
shown that for an present typical parameters of JCQ $E_{J}/E_{ch}\sim 0.02$,
the loss of the fidelity per elementary operation is about $10^{-4}$ which
satisfy the DiVincenzo's low decoherence criterion. By appropriately
improving the design of the Josephson junction, namely, decreasing $%
E_{J}/E_{ch}$ to $\sim 0.01,$ the loss of fidelity per elementary operation
can decrease to $10^{-5}$ even smaller$.$


\end{abstract}

\section{Introduction}

A quantum computer can perform certain tasks which no classical one is able
to do in acceptable times \cite{qcomputer}. A quantum bit (qubit) is a
quantum system with two levels which will be a cell for storing and
processing information in a future quantum computer. So it is a vital task
to find out the physical realizations of the qubits. In last years,
physicists have proposed several qubit models which are based on ion traps 
\cite{Ion-trap}, QED systems \cite{QED}, nuclear spins of large numbers of
identical molecules \cite{nuclear-spin}, quantum dots \cite{qdots},
Josephson junction \cite{Josephson} and so on. Because solid state qubits
can be embedded in electronic circuits as well as scaled up to a large
numbers, they are taken as a particularly promising candidates \cite%
{promising} of qubits for quantum computation. The Josephson junction qubit
is one of this kinds of models.

Decoherence, one of the most difficult problems to be dealt with in quantum
computation exists in all of the qubit models. In general, decoherence comes
from the interaction of the qubits and their environment. But for some qubit
models, for example, quantum dots and Josephson junctions, the decoherence
also results from intrinsic source of error, such as quantum leakage \cite%
{leakage01} \cite{leakage02}. The quantum leakage is this kind of process
that the system working in the computational Hilbert space leaks out to
higher states. The leakage exists in many quantum processes \cite{leakage03} 
\cite{Sun}. It is attracted a particular attention in the implementations of
qubits and quantum gates for quantum computation \cite{leakage01} because
they must satisfy the DiVincenzo's checklist five criteria \cite{DiVincenzo}
one of which is low decoherence (so that error correction techniques may be
used in a fault-tolerant manner)---an approximate benchmark is the loss of
fidelity no more than $10^{-4}$ per elementary quantum gate operation.

In \cite{leakage01}, Fazio \textit{et al}. investigated the leakage and
fidelity of the Josephson charge qubit (JCQ) operations. Where the
eigenvalues and eigenstates of the JCQ Hamiltonian are obtained through
diagonalizing the Hamiltonian. In fact, they can be obtained in virtue of
the perturbation theory of quantum mechanics \cite{quant-ph/0112026}. In
particular, as pointed in \cite{leakage01} the eigenvalues and eigenstates
can be obtained by solving the eigen-equation of the JCQ Hamiltonian which
in fact is the Mathieu equation \cite{Mathieufun}. The eigenvalues and
eigenstates of the JCQ Hamiltonian correspond to the characteristic values
and characteristic states of the Mathieu equation. In this paper we will use
the well researched Mathieu functions investigating the loss of fidelity for
the single JCQ.

\section{Dynamics of the JCQ}

The single JCQ was first introduced by Shnirman \emph{et al.} \cite%
{Josephson}. Since then, much interest has been attracted into this topic.
The simplest JCQ can be designed as Fig.1 (refer to Fig.1 of Ref. \cite%
{promising}). \FRAME{ftbpFU}{104.2564pt}{144.9796pt}{0pt}{\Qcb{A Josephson
charge qubit in its simplest design formed by a superconducting
single-charge box.}}{}{Figure}{\special{language "Scientific Word";type
"GRAPHIC";display "USEDEF";valid_file "T";width 104.2564pt;height
144.9796pt;depth 0pt;original-width 94.0911pt;original-height
138.8433pt;cropleft "0";croptop "1";cropright "1";cropbottom
"0";tempfilename 'I07F1201.wmf';tempfile-properties "XPR";}}Ignoring the
resistance in the circuit, from the Josephson relations%
\begin{equation}
I=I_{c}\sin \varphi ,\text{ }\frac{\Phi _{0}}{2\pi }\frac{d\varphi }{dt}%
=U_{J},  \label{e1}
\end{equation}%
and the current conservation we have 
\begin{equation}
C_{J}\frac{\Phi _{0}}{2\pi }\frac{d^{2}\varphi }{dt^{2}}+I_{c}\sin \varphi
=C_{g}\dot{U}_{g},  \label{e2}
\end{equation}%
where $I_{c}$ is the critical current of the Josephson junction, $\Phi
_{0}=\pi \hbar /e$ is a magnetic flux quantum, $\varphi $ is the gauge
invariant phase of the superconducting junction, $U_{J}$ is the voltage of
the Josephson junction and $U_{g}$ is the gate voltage. Due to $%
U_{J}+U_{g}=V_{g}$, Eq.(\ref{e2}) can be written as%
\begin{equation}
C\left( \frac{\Phi _{0}}{2\pi }\right) ^{2}\frac{d^{2}\varphi }{dt^{2}}+%
\frac{\Phi _{0}I_{c}}{2\pi }\sin \varphi =0,  \label{e3}
\end{equation}%
where $C=\left( C_{J}+C_{g}\right) .$ Thus, we can obtain the Lagrangian of
the JCQ as 
\begin{equation}
\mathcal{L}=\frac{1}{2}C\left( \frac{\Phi _{0}}{2\pi }\right) ^{2}\dot{%
\varphi}^{2}+\frac{\Phi _{0}I_{c}}{2\pi }\cos \varphi .  \label{e4}
\end{equation}%
The Euler-Lagrange equation can be used to check that the Lagrangian
produces the correct classical equations of motion,%
\begin{equation}
\frac{d}{dt}\frac{\partial \mathcal{L}}{\partial \dot{\varphi}}-\frac{%
\partial \mathcal{L}}{\partial \varphi }=0.  \label{e5}
\end{equation}%
By definition, the conjugate variable to $\varphi $ is,%
\begin{equation}
p=\frac{\partial \mathcal{L}}{\partial \dot{\varphi}}=C\left( \frac{\Phi _{0}%
}{2\pi }\right) ^{2}\dot{\varphi}.  \label{e6}
\end{equation}%
On the other hand, the charge on the Josephson junction capacitance and the
gate capacitance are 
\begin{eqnarray}
q &=&C_{J}U_{J}=C_{J}\frac{\Phi _{0}}{2\pi }\dot{\varphi}\equiv n2e,  \notag
\\
q_{g} &=&C_{g}U_{g}=-C_{g}\frac{\Phi _{0}}{2\pi }\dot{\varphi}\equiv n_{g}2e,
\label{e7}
\end{eqnarray}%
where $n$ is the Cooper pairs pass through the Josephson junction, and $%
n_{g} $ is the number of two-unit charge $2e$ on the gate capacitance. From
Eqs.(\ref{e6}) and (\ref{e7}) we have 
\begin{eqnarray}
p &=&\frac{\Phi _{0}}{2\pi }\left( q-q_{g}\right) =C\left( \frac{\Phi _{0}}{%
2\pi }\right) ^{2}\dot{\varphi}  \notag \\
&=&\frac{\Phi _{0}}{2\pi }2e\left( n-n_{g}\right) .  \label{e8}
\end{eqnarray}%
So 
\begin{equation}
\dot{\varphi}=\frac{2\pi }{\Phi _{0}}\frac{2e}{C}\left( n-n_{g}\right) .
\label{e9}
\end{equation}%
Now we can construct the Hamiltonian via Legendre transformation, 
\begin{eqnarray}
\mathcal{H} &=&\dot{\varphi}p-\mathcal{L}=C\left( \frac{\Phi _{0}}{2\pi }%
\right) ^{2}\dot{\varphi}^{2}-\frac{\Phi _{0}I_{c}}{2\pi }\cos \varphi 
\notag \\
&=&\frac{2e^{2}}{C}\left( n-n_{g}\right) ^{2}-\frac{\Phi _{0}I_{c}}{2\pi }%
\cos \varphi .  \label{e10}
\end{eqnarray}%
Setting $E_{ch}=\frac{2e^{2}}{C},$ $E_{J}=\frac{\Phi _{0}I_{c}}{2\pi },$ we
have%
\begin{equation}
\mathcal{H}=E_{ch}\left( n-n_{g}\right) ^{2}-E_{J}\cos \varphi .  \label{e11}
\end{equation}%
So we have $\left[ p,\varphi \right] =-i\hbar ,$ and $\left[ n,\varphi %
\right] =-i\hbar \pi /\Phi _{0}e.$ As Ref. \cite{Josephson}, to shorten
notations we use units where $e=1,$ $\hbar =1.$ So $\left[ n,\varphi \right]
=-i.$ Due to the conjugate relationship we have $n=-i\frac{d}{d\varphi }$.
Therefore the Hamiltonian becomes%
\begin{equation}
\mathcal{H}=E_{ch}\left( -\frac{d^{2}}{d\varphi ^{2}}+2n_{g}i\frac{d}{%
d\varphi }+n_{g}^{2}\right) -E_{J}\cos \varphi .  \label{e12}
\end{equation}%
Setting the eigenstates of $\mathcal{H}$ be $\Psi _{n}$ and according to the
eigen-equation, we have 
\begin{equation}
\frac{d^{2}\Psi _{n}}{d\varphi ^{2}}+2n_{g}i\frac{d\Psi _{n}}{d\varphi }%
-\left( n_{g}^{2}-\frac{E_{n}}{E_{ch}}-\frac{E_{J}\cos \varphi }{E_{ch}}%
\right) \Psi _{n}=0.  \label{e13}
\end{equation}%
Setting $\Psi _{n}=e^{ix\varphi }\psi _{n}\left( \varphi \right) ,$ and $%
x=1/2,$ we have%
\begin{equation}
\frac{d^{2}\psi _{n}}{d\varphi ^{2}}+2ik\frac{d\psi _{n}}{d\varphi }-\left(
k^{2}-\frac{E_{n}}{E_{ch}}-\frac{E_{J}\cos \varphi }{E_{ch}}\right) \psi
_{n}=0,  \label{e14}
\end{equation}%
where $k=\frac{1}{2}-n_{x}.$ If we modulate the controllable gate voltage $%
V_{g}$ and make $n_{g}=x=\frac{1}{2},$ then Eq.(\ref{e14}) become%
\begin{equation}
\frac{d^{2}\psi _{n}}{d\varphi ^{2}}+\left( \lambda -2v\cos \varphi \right)
\psi _{n}=0,  \label{e15}
\end{equation}%
where%
\begin{equation}
\lambda =\frac{E_{n}}{E_{ch}},\text{ }v=-\frac{E_{J}}{2E_{ch}}.  \label{e16}
\end{equation}

Eq.(\ref{e15}) is the canonical form of the Mathieu equation \cite%
{Mathieufun}, its characteristic functions called Mathieu functions. The
Mathieu functions were introduced by Mathieu \cite{Mathieu01} when analyzing
the movements of membranes of elliptical shape. Since then the characters of
the Mathieu functions have been investigated by Mathieu and others \cite%
{Mathieufun}. In recent years, the functions have been attracted much
attention because they have some applications in many fields of physics \cite%
{Mathieu02}. The Mathieu equation has the well known periodic solutions%
\begin{eqnarray*}
&&\left\{ 
\begin{array}{c}
ce_{2n}\left( \varphi ,v\right) \text{ \ even solutions with period }\pi \\ 
\text{ \ \ \ \ \ \ with eigenvalues }a_{2n}\left( v\right) , \\ 
se_{2n+2}\left( \varphi ,v\right) \text{ odd solutions with period }\pi \\ 
\text{ \ \ \ \ \ \ \ \ with eigenvalues }b_{2n+2}\left( v\right) .%
\end{array}%
\right. \\
&& \\
&&\left\{ 
\begin{array}{c}
ce_{2n+1}\left( \varphi ,v\right) \text{ even solutions with period }2\pi \\ 
\text{ \ \ \ \ \ \ \ with eigenvalues }a_{2n+1}\left( v\right) , \\ 
se_{2n+1}\left( \varphi ,v\right) \text{ odd solutions with period }2\pi \\ 
\text{ \ \ \ \ \ \ \ with eigenvalues }b_{2n+1}\left( v\right) .%
\end{array}%
\right.
\end{eqnarray*}%
It has been pointed that the periodic boundary conditions $\psi _{n}\left(
\varphi =0\right) =\psi _{n}\left( \varphi =\pi \right) $ singles out only
the $2\pi $ periodic Mathieu eigenfunctions $ce_{2n},se_{2n}$ for an integer 
$x$ and the $\pi $-anti-periodic Mathieu eigenfunctions $ce_{2n+1},se_{2n+1}$
\cite{Mathieu03} for a half-integer $x$. So when one suddenly switch the
offset charge from idle point to the degeneracy point $n_{g}=\frac{1}{2}$
(or another half-integer), we can obtain the eigenvalues (leave over the
first five terms) and the eigenfunctions (leave over the first three terms)
of the Hamiltonian $\mathcal{H}$ as%
\begin{eqnarray}
E_{1}^{e} &=&E_{ch}\left( 1+v-\frac{v^{2}}{8}-\frac{v^{3}}{64}-\frac{v^{4}}{%
1536}-\cdot \cdot \cdot \right) ,  \notag \\
E_{1}^{o} &=&E_{ch}\left( 1-v-\frac{v^{2}}{8}+\frac{v^{3}}{64}-\frac{v^{4}}{%
1536}+\cdot \cdot \cdot \right) ;  \label{e17}
\end{eqnarray}%
\begin{eqnarray}
\Psi _{1}^{e} &=&e^{i\varphi /2}\psi _{1}^{e}=e^{i\varphi /2}\left[ \cos
\varphi -\frac{v}{8}\cos 3\varphi \right.  \notag \\
&&\left. +v^{2}\left( \frac{\cos 5\varphi }{192}-\frac{\cos 3\varphi }{64}-%
\frac{\cos \varphi }{128}\right) +\cdot \cdot \cdot \right] ,  \notag \\
\Psi _{1}^{o} &=&e^{i\varphi /2}\psi _{1}^{o}=e^{i\varphi /2}\left[ \sin
\varphi -\frac{v}{8}\sin 3\varphi \right.  \notag \\
&&\left. +v^{2}\left( \frac{\sin 5\varphi }{192}+\frac{\sin 3\varphi }{64}-%
\frac{\sin \varphi }{128}\right) +\cdot \cdot \cdot \right] .  \label{e18}
\end{eqnarray}%
Enlightened by \cite{PRA66-012312}, we have%
\begin{eqnarray}
\left\vert \psi _{1}^{e}\right\rangle &=&\sqrt{\frac{32}{64+v^{2}+\delta }}%
\times  \notag \\
&&\left[ \left\vert 0\right\rangle +\left\vert 1\right\rangle +\frac{v}{8}%
\left( \left\vert -1\right\rangle +\left\vert 2\right\rangle \right) +\cdot
\cdot \cdot \right] ,  \notag \\
\left\vert \psi _{1}^{o}\right\rangle &=&\sqrt{\frac{32}{64+v^{2}+\delta }}%
\times  \notag \\
&&\left[ \left\vert 0\right\rangle -\left\vert 1\right\rangle +\frac{v}{8}%
\left( \left\vert -1\right\rangle -\left\vert 2\right\rangle \right) +\cdot
\cdot \cdot \right] .  \label{e19}
\end{eqnarray}%
Here, $\delta $ denotes the higher-order effects of $v^{2}.$ When $E_{J}\ll
E_{ch},$ we can set $\delta \rightarrow 0.$ So after a time $t,$ the initial
state $\left\vert \beta _{0}\right\rangle =\cos \theta \left\vert
0\right\rangle +\sin \theta \left\vert 1\right\rangle $ in the system
becomes 
\begin{eqnarray}
\left\vert \Psi \right\rangle _{R} &=&U_{R}\left\vert \beta _{0}\right\rangle
\notag \\
&=&\sum_{j=e,o}e^{-iE_{1}^{j}t}\left\vert \psi _{1}^{i}\right\rangle
\left\langle \psi _{1}^{i}\right\vert \left. \beta _{0}\right\rangle  \notag
\\
&=&e^{-iE_{1}^{e}t}\left\vert \psi _{1}^{e}\right\rangle \left\langle \psi
_{1}^{e}\right\vert \left. \beta _{0}\right\rangle
+e^{-iE_{1}^{o}t}\left\vert \psi _{1}^{o}\right\rangle \left\langle \psi
_{1}^{o}\right\vert \left. \beta _{0}\right\rangle .  \label{e20}
\end{eqnarray}%
It is shown that by using the Mathieu functions we can obtain more exact
results of the eigenvalues and eigenstates of $\mathcal{H}$ than previous
researches$.$ In particular, the eigenvalues and eigenstates can approximate
to a arbitrary higher order of the Mathieu functions can be obtained.

On the other hand, because the Josephson energy $E_{J}$ is much smaller than
the charging energy $E_{ch}$, and both of them are smaller than the
superconducting energy gap $\Delta $, the Hamiltonian Eq.(\ref{e11}) can be
parameterized by the number of the Cooper pairs $n$ through the junction as%
\begin{eqnarray}
\mathcal{H} &=&\sum_{n}\left\{ E_{ch}\left( n-n_{g}\right) ^{2}\left\vert
n\right\rangle \left\langle n\right\vert \right.  \notag \\
&&\left. -\frac{1}{2}E_{J}\left[ \left\vert n\right\rangle \left\langle
n+1\right\vert +\left\vert n+1\right\rangle \left\langle n\right\vert \right]
\right\} .  \label{e21}
\end{eqnarray}%
When $n_{g}$ is modulated to a half-integer, say $n_{g}=1/2$ and the
charging energies of two adjacent states are closed each other, the
Josephson tunneling mixes them strongly. Then, the system can be reduces to
a two-state system (qubit) because all other charge states have much higher
energy and they can be neglected, the Hamiltonian is $\emph{approximately}$
reads%
\begin{equation}
\mathcal{H}_{I}=E_{ch}\left( n-\frac{1}{2}\right) ^{2}\sigma _{z}-\frac{1}{2}%
E_{J}\sigma _{x}.  \label{e22}
\end{equation}%
This is an ideal Hamiltonian of the qubit. By choosing the reference point
of the energy at $E_{0}=E_{ch}/4,$ the Hamiltonian can deduce to%
\begin{equation}
\mathcal{H}_{I}=-\frac{E_{J}}{2}\sigma _{x},  \label{e23}
\end{equation}%
which has the eigenvalues and eigenstates as%
\begin{eqnarray}
E_{0} &=&-\frac{E_{J}}{2},\text{ }\left\vert \varphi _{0}\right\rangle =%
\frac{1}{\sqrt{2}}\left( \left\vert 0\right\rangle +\left\vert
1\right\rangle \right) ,  \notag \\
E_{1} &=&\frac{E_{J}}{2},\text{ }\left\vert \varphi _{1}\right\rangle =\frac{%
1}{\sqrt{2}}\left( \left\vert 0\right\rangle -\left\vert 1\right\rangle
\right) .  \label{e24}
\end{eqnarray}%
So after a time $t$ the initial state $\left\vert \beta _{0}\right\rangle $
becomes%
\begin{eqnarray}
\left\vert \Psi \right\rangle _{I} &=&U_{I}\left\vert \beta
_{0}\right\rangle =\sum_{i=1,2}e^{-iE_{i}t}\left\vert \varphi
_{i}\right\rangle \left\langle \varphi _{i}\right\vert \left. \beta
_{0}\right\rangle  \notag \\
&&\text{ \ \ \ \ \ }%
\begin{tabular}{l}
$=e^{iE_{J}t/2}\left\vert \varphi _{0}\right\rangle \left\langle \varphi
_{0}\right\vert \left. \beta _{0}\right\rangle $%
\end{tabular}
\notag \\
&&\text{ \ \ \ \ \ }%
\begin{tabular}{l}
$+e^{-iE_{J}t/2}\left\vert \varphi _{1}\right\rangle \left\langle \varphi
_{1}\right\vert \left. \beta _{0}\right\rangle .$%
\end{tabular}
\label{e25}
\end{eqnarray}%
It is shown that $\left\vert \Psi \right\rangle _{R}$ and $\left\vert \Psi
\right\rangle _{I}$ are different. The difference derives from the quantum
leakage. In the following, we shall investigate the loss of fidelity due to
the quantum leakage for the JCQ.

\section{Leakage and Fidelity of JCQ}

The leakage of the JCQ is in fact the probability of initial state $%
\left\vert \beta _{0}\right\rangle $ leaks out to higher states after some
time in the practical system. It can be defined as%
\begin{eqnarray}
L &=&\max \sum_{i\neq 0,1}\text{ }_{R}\left\langle \Psi \right\vert \Pi
_{i}\left\vert \Psi \right\rangle _{R}  \notag \\
&=&1-\min \sum_{i=0,1}\text{ }_{R}\left\langle \Psi \right\vert \Pi
_{i}\left\vert \Psi \right\rangle _{R}  \label{e26}
\end{eqnarray}%
where $\Pi _{0}=\left\vert 0\right\rangle \left\langle 0\right\vert ,$ $\Pi
_{1}=\left\vert 1\right\rangle \left\langle 1\right\vert ,$ $\cdot \cdot
\cdot ,$ $\Pi _{i}=\left\vert i\right\rangle \left\langle i\right\vert $ are
project operators$;$ $\left\vert \Psi \right\rangle _{R}$ is the finally
state. The loss of fidelity is the probability by measuring the state $%
\left\vert \Psi \right\rangle _{R}$ with the project operators BUT $\Pi
_{0}=\left\vert 0\right\rangle \left\langle 0\right\vert $ and $\Pi
_{1}=\left\vert 1\right\rangle \left\langle 1\right\vert .$ By use of Eq.(%
\ref{e20}) we have%
\begin{eqnarray}
\sum_{i=0,1}\text{ }_{R}\left\langle \Psi \right\vert \Pi _{i}\left\vert
\Psi \right\rangle _{R} &=&\sum_{i=0,1}\left\langle \beta _{0}\right\vert
U_{R}^{\dagger }\Pi _{i}U_{R}\left\vert \beta _{0}\right\rangle  \notag \\
&&%
\begin{tabular}{l}
$=\left\langle \psi _{1}^{e}\right\vert \left. \psi _{1}^{e}\right\rangle
\left\vert \left\langle \beta _{0}\right\vert \left. \psi
_{1}^{e}\right\rangle \right\vert ^{2}$%
\end{tabular}
\notag \\
&&%
\begin{tabular}{l}
$+\left\langle \psi _{1}^{o}\right\vert \left. \psi _{1}^{o}\right\rangle
\left\vert \left\langle \beta _{0}\right\vert \left. \psi
_{1}^{o}\right\rangle \right\vert ^{2}$.%
\end{tabular}
\label{e27}
\end{eqnarray}%
Because of%
\begin{eqnarray}
\left\langle \psi _{1}^{e}\right\vert \left. \beta _{0}\right\rangle &=&%
\sqrt{\frac{32}{64+v^{2}+\delta }}\left( \cos \theta +\sin \theta \right) , 
\notag \\
\left\langle \psi _{1}^{o}\right\vert \left. \beta _{0}\right\rangle &=&%
\sqrt{\frac{32}{64+v^{2}+\delta }}\left( \cos \theta -\sin \theta \right) , 
\notag \\
\left\langle \psi _{1}^{o}\right\vert \left. \psi _{1}^{o}\right\rangle
&=&\left\langle \psi _{1}^{e}\right\vert \left. \psi _{1}^{e}\right\rangle =%
\frac{64}{64+v^{2}+\delta },  \notag \\
\left\langle \psi _{1}^{e}\right\vert \left. \psi _{1}^{o}\right\rangle &=&0,
\label{e28}
\end{eqnarray}%
we have%
\begin{equation}
L=1-\left( \frac{64}{64+v^{2}+\delta }\right) ^{2}\approx 1-\left( \frac{64}{%
64+v^{2}}\right) ^{2}.  \label{e29}
\end{equation}%
On the other hand, in general, the fidelity is defined as $F\left( \rho
,\sigma \right) =tr\sqrt{\rho ^{\frac{1}{2}}\sigma \rho ^{\frac{1}{2}}}$ for
two arbitrary states $\rho $ and $\sigma ,$ and $F\left( \left\vert \psi
\right\rangle ,\left\vert \varphi \right\rangle \right) =\left\langle
\varphi \right\vert \left. \psi \right\rangle $ for two pure states $%
\left\vert \psi \right\rangle $ and $\left\vert \varphi \right\rangle $ \cite%
{Nielson-Chuang}$.$ By using $\left\vert \Psi \right\rangle _{I}$ and $%
\left\vert \Psi \right\rangle _{R}$ we can straightforwardly calculate the
loss of fidelity, it is also%
\begin{equation}
L=1-\left\vert _{I}\left\langle \Psi \right\vert \Pi \left\vert \Psi
\right\rangle _{R}\right\vert ^{2}\approx 1-\left( \frac{64}{64+v^{2}}%
\right) ^{2}.  \label{e30}
\end{equation}%
It means that if we do not consider the interaction of the qubit and the
environment, the loss of fidelity is just the quantum leakage $L$. So the
relation of the fidelity and leakage is $F=1-L.$

For present typical parameters of Josephson junction $E_{J}/E_{ch}=0.02$ 
\cite{leakage01}$,$ we can obtain the fidelity $F=0.9999875000$ (calculated
by using our formula)$,$ which is agreement with $F=0.9999823223$
(calculated by using the Eq.(7) of Ref.\cite{leakage01}) very well. It can
be easily seen that the loss of fidelity due to the leakage will be
decreased by an appropriate choice of the device parameters. For example,
the fidelity will increase to $F=0.9999968750$ for $E_{J}/E_{ch}=0.01,$ and
to $F=0.9999992188$ for $E_{J}/E_{ch}=0.005,$ which shows that the loss of
the fidelity is about $10^{-6}$ per elementary gate operation as $%
E_{J}/E_{ch}\sim 0.005$. According to DiVincenzo's low decoherence criterion
the loss of fidelity is tolerable. From the subsection II we know $E_{ch}=%
\frac{2e^{2}}{C},$ $E_{J}=\frac{\Phi _{0}I_{c}}{2\pi }$, and we use units
where $e=1,$ $\hbar =1.$ So $\Phi _{0}=\pi \hbar /e=\pi $ in the units.
Ttherefore $E_{ch}=\frac{2}{C},$ $E_{J}=\frac{I_{c}}{2}$. To decrease $%
E_{J}/E_{ch}$ one should decrease the critical current $I_{c}$ OR the total
capacitance $C$. However, in Ref.\cite{Liang} we investigate the short-time
decoherence of the JCQ, where the increasing of the critical current $I_{c}$
AND the decreasing of the total capacitance $C$ are needed for decreasing
the decoherence derived from the interaction of the system and its
environment. From the analysis of the two papers we know that to decrease
the decoherence not only from quantum leakage but from the environment one
may improve the design of the JCQ through increasing the critical current $%
I_{c}$ AND decreasing the total capacitance $C$. But the design should keep $%
E_{J}/E_{ch}\leq 0.005\sim 0.02.$

\section{Conclusions}

In this paper we investigated the loss of fidelity due to the quantum
leakage for JCQ. Our researches are based on a lot of results of Mathieu
functions. It is shown that our results agree with previous corresponding
investigations very well. However, our work can expand to higher order
approximation easily because of the well researched Mathieu functions. In
particular, our results provide a feedback on how to improve the design of
the JCQ for quantum computation. It is shown that decreasing the critical
current and decreasing the Josephson capacitance and gate capacitance can
decrease the decoherence from the quantum leakage. However, in order to
decrease the total decoherence one may improve the design by increasing the
critical current $I_{c}$ AND decreasing the total capacitance $C$. But the
design should keep $E_{J}/E_{ch}\leq 0.005\sim 0.02.$ So we think that it is
necessary to develop the technology of increasing the Josephson critical
current and decreasing the capacitances in the small Josephson junctions in
order to make the JCQ suitable for quantum computation.

\begin{acknowledgement}
The author (X.T.L) would like to thank Rosario Fazio for helpful discussions
and referees for their constructive advice. This work were supported by the
National Natural Science Foundation of China (NSFC), grant numbers 10347133,
10347134 and Ningbo City Youth Foundation, grant numbers 2004A620003,
2003A620005.
\end{acknowledgement}

\end{document}